\documentclass{emulateapj}

\begin{document}

\title{A \emph{Spitzer} Study of Dusty Disks in the Scorpius-Centaurus
    OB Association}
\author{C.\ H.\ Chen\altaffilmark{1}, M. \ Jura\altaffilmark{2},
        K. D. Gordon\altaffilmark{3}, M. Blaylock\altaffilmark{3}}
\altaffiltext{1}{National Research Council Resident Research Associate,
    Jet Propulsion Laboratory, M/S 169-506, California Institute of 
    Technology, 4800 Oak Grove Drive, Pasadena, CA 91109; 
    christine.chen@jpl.nasa.gov}
\altaffiltext{2}{Department of Physics and Astronomy, University of 
    California, Los Angeles, CA 90095-1562; jura@astro.ucla.edu}
\altaffiltext{3}{Steward Observatory, University of Arizona,
    Tucson, AZ 85721}

\begin{abstract}
We have obtained \emph{Spitzer Space Telescope} MIPS observations of 40 F- 
and G-type common proper motion members of the Scorpius-Centaurus OB 
Association with ages between 5 and 20 Myr at 24 $\mu$m and 70 $\mu$m.
We report the detection of fourteen objects which possess 24 $\mu$m fluxes 
$\geq$30\% larger than their predicted photospheres, tentatively 
corresponding to a disk fraction of $\geq$35\%, including seven objects which 
also possess 70 $\mu$m excesses $\geq$100$\times$ larger than their predicted 
photospheres. The 24 $\mu$m plus 70 $\mu$m excess sources possess high 
fractional infrared luminosities, $L_{IR}/L_{*}$ = 7 $\times$ 10$^{-4}$ - 3 
$\times$ 10$^{-3}$; either they possess optically thin, dusty $\beta$ 
Pictoris-like disks or compact, opaque HD 98800-like disks.
\end{abstract}


\section{Introduction}
Circumstellar disks around young stars with ages $<$100 Myr are the likely 
birth places of planets and smaller bodies, analogous to asteroids, comets, 
and Kuiper-belt objects in our Solar System. However, how optically thick 
disks with interstellar gas-to-dust ratios evolve into Solar Systems is not 
well understood. Recent studies of the 10 Myr old TW Hydrae Association 
(Weinberger et al. 2004) and the 30 Myr Tucana-Horologium Association 
(Mamajek et al. 2004) find that warm circumstellar dust ($T_{gr}$ = 200 - 
300 K), if present, reradiates less than 0.1\% - 0.7\% of the stellar 
luminosity, $L_{IR}/L_{*}$ $<$ 1 - 7 $\times$ 10$^{-3}$. If the majority of 
stars possess Solar Systems, why do some stars, such as TW Hydrae, Hen 3-600, 
HD 98800 and HR 4796A in the TW Hydrae Association, still possess large 
quantities of gas and/or dust? In addition, the age $\sim$10 Myr appears to 
mark a transition in the gross properties of observed disks. Both optically 
thick, gaseous disks and optically thin, dusty disks have been discovered 
around 10 Myr old stars. Pre-main sequence T-Tauri and Herbig Ae/Be stars 
with ages $<$5 Myr typically possess both gas and dust while debris disks 
with ages $>$100 Myr are typically dusty and gas poor.

Detailed studies of the gas and dust content of 10 Myr old circumstellar 
disks suffer from small number statistics. Circumstellar disks around
main sequence stars were discovered using the \emph{IRAS} satellite
(Backman \& Paresce 1993); however, the sensitivity of \emph{IRAS} limited 
studies of circumstellar disks to dust around T-Tauri stars, Herbig Ae/Be 
stars, and early-type main sequence stars. The excellent sensitivity of the 
\emph{Spitzer Space Telescope} allows us not only to characterize better
already known systems but also to identify previously unknown circumstellar 
disk systems around more distant objects and around fainter, later-type stars.
We have obtained \emph{Spitzer Space Telescope} MIPS observations of F- 
and G-type common proper motion members of the Scorpius-Centaurus OB 
Association with ages between 5 and 20 Myr to search for dusty circumstellar 
disks which may be transitioning from optically thick, gaseous systems to 
optically thin, dusty systems in order to understand better the evolution of 
gas and dust around young stars.

The Scorpius-Centaurus OB association, with typical stellar distances of 118 
- 145 pc, is the closest OB assoication to the Sun and contains three 
subgroups: Upper Scorpius, Upper Centaurus Lupus, and Lower Centaurus Crux, 
with estimated ages of $\sim$5 Myr (Preibisch et al. 2002), 17 Myr, and 
16 Myr (Mamajek, Meyer, \& Liebert 2002) respectively. Member stars with 
spectral type F and earlier have been identified via common proper motion 
studies using \emph{Hipparcos} data (de Zeeuw et al. 1999) while later-type 
members have been identified via indicators of youth (i.e. high coronal
x-ray activity and large lithium abundance: Preibisch et al. 2002; Mamajek et
al. 2002). The close proximity of Scorpius-Centaurus and the youth of its 
consituent stars make this association an excellent laboratory for studying
the formation and evolution of planetary systems. A recent survey of 5 
members of Upper Scorpius detected $N$-band excesses associated with two 
stars: J161411.0-230536 and HD 143006 (Mamajek et al. 2004).

We report a first set of results from a \emph{Spitzer} search for dusty disks 
around 130 F-, G-, and K-type members of nearby (within 150 pc of the Sun) 
OB Associations using MIPS at 24 $\mu$m and 70 $\mu$m. We have obtained 
24 $\mu$m and 70 $\mu$m photometry for 41 F- and G-type stars, including 5 
members of Upper Scorpius, 11 members of Upper Centaurus Crux, and 25 members 
of Lower Centaurus Crux, whose membership has been determined via 
\emph{Hipparcos} proper motion studies (de Zeeuw et al. 1999). Using Monte 
Carlo simulations, De Zeeuw et al. (1999) estimate that 32\% of their 
proposed Lower Centaurus Crux members and 24\% of their proposed Upper 
Centaurus Lupus members may be interlopers and not true association members 
because the stellar radial velocities are not measured. We cross-correlated 
our target list with published lists of confirmed and rejected 
Scorpius-Centaurus group members (Mamjek et al. 2002). One star (HD 113376) 
is excluded as a member based upon the lack of youth indicators (H$\alpha$ 
emission, \emph{ROSAT} flux, lithium abundance). Four others (HD 105070, HD 
108568, HD 108611, HD 119022, HD 138009) are confirmed as members from their 
subgiant surface gravities, high lithium abundances, strong x-ray emission 
($L_{x}$ = $10^{29}$ - $10^{30}$ ergs s$^{-1}$), and proper motions 
consistent with the high mass members. We list the targets along with their 
spectral types, distances, $V$-band extinctions (Sartori et al. 2003), 
\emph{ROSAT} fluxes, and OB Association subgroup memberships in Table 1.

\section{Observations}
Our data were obtained  using the Multiband Imaging Spectrometer for 
\emph{Spitzer} (MIPS) (Rieke et al. 2004) on the \emph{Spitzer Space 
Telescope} (Werner et al. 2004) in photometry mode at 24 $\mu$m and 
70 $\mu$m (default scale). Each of our targets was observed either during
20 - 26 February 2004 or during 14 - 22 March 2004 using integration times of 
48.2 sec at 24 $\mu$m and 93 sec at 70 $\mu$m. The data were reduced and 
combined using the Data Analysis Tool (DAT) version 2.80 developed by the 
MIPS instrument team (Gordon et al. 2004). Extra processing steps beyond those 
in the DAT were applied to remove known transient effects associated with the 
70 micron detectors to achieve the best possible sensitivity. While 
point sources are well calibrated using the stim flashes, extended sources 
(eg., the background) show small responsivity drifts with respect to the 
point source calibration. As a result, the background in uncorrected mosaics 
displays significant structure associated with detector columns. This 
detector dependent structure is removed by subtracting column averages from 
each exposure with the source region masked. In addition, a pixel dependent 
time filter is applied (with the source region masked) to remove small pixel 
dependent residuals. These corrected images are then combined to produce the 
final mosaic used for the source detection.

The estimated 70 $\mu$m sky backgrounds, extrapolated from COBE data, 
using the IRSKY Batch Inquiry System (IBIS), suggest that our fields-of-view
possess medium to high 70 $\mu$m cirrus backgrounds (see Table 2). To
avoid confusion with cirrus clumps, we visually searched for bright sources 
in each of our 70 $\mu$m fields-of-view. If no object was detected at the
center of the array, we placed detection limits on the fluxes, computed 
from the sky noise measured in a background annulus. This sky noise 
includes detector noise and noise due to cirrus structures present in 
the image. Since these observations were taken in a region with high cirrus, 
the detection limits are dominated by cirrus noise. In fact, examination of 
the 70 micron images reveals significant cirrus structures which make our 
simple aperture photometry detection limits conservative upper limits. 

We used aperture photometry to measure the fluxes, by finding the average 
brightness of a pixel in a ``sky'' annulus around the source, subtracting 
this value from each pixel in the aperture, and then summing the flux 
in the aperture. Since our observations are diffraction limited and the pixel 
scale for the 24 $\mu$m and 70 $\mu$m detectors are different, we use 
different aperture radii and sky annuli for the 24 $\mu$m and 70 $\mu$m data. 
We used 15$\arcsec$ (6 pixel) and a 29.5$\arcsec$ (3 pixel) radius apertures 
to estimate the source flux and 30$\arcsec$ (12 pixel) to 43$\arcsec$ 
(17 pixel) and 40$\arcsec$ (4 pixel) to 80$\arcsec$ (8 pixel) sky annuli to 
estimate the sky backgrounds at 24 $\mu$m and 70 $\mu$m, respectively. 
These apertures are not large enough to contain all of the photons from a 
diffraction limited point source; therefore, we applied scalar aperture 
corrections of 1.147 and 1.267 at 24 $\mu$m and 70$\mu$m, respectively, 
inferred from \emph{Spitzer} Tiny Tim models of the Point Spread Function 
(Krist 2002), to extrapolate the object fluxes from the photon fluxes in the 
apertures. We flux calibrated our data assuming conversion factors of 1.042 
$\mu$Jy arcsec$^{-2}$/(DN sec$^{-1}$) at 24 $\mu$m and 1.58$\times$10$^{4}$ 
$\mu$Jy arcsec$^{-2}$/(DN sec$^{-1}$) at 70 $\mu$m. We list the 
uncolor-corrected observed 24 $\mu$m and 70 $\mu$m fluxes in Table 2. Current 
observations of standard stars suggest that MIPS 24 $\mu$m photometry has an 
uncertainty of $\sim$10\% for stars fainter than 4 Jy and that MIPS 70 $\mu$m 
photometry has an uncertainty of $\sim$20\% for objects brighter than 50 mJy.

We modeled the stellar photospheres of our objects by minimum $\chi^{2}$ 
fitting extinction corrected published photometry from the literature, using 
fluxes at wavelengths shorter than 3 $\mu$m, to 1993 Kurucz stellar 
atmospheres using the package Best Kurucz developed by the MIPS instrument 
team (Su et al. 2004). We input measured extinctions, $A_{V}$, stellar 
effective temperatures, $T_{*}$, and gravities, $\log g$, from Sartori et al. 
(2003), extinction-corrected the fluxes using the Cardelli, Clayton, \& 
Mathis (1989) extinction law, searched a grid of parameter space around our 
assumed values, and found an atmosphere model (effective temperature, 
metallicity, gravity) which fit the data the best. Where possible, we 
included fluxes from TD 1 (Thompson et al. 1978), \emph{Hipparcos}, the 
General Catalog of Photometric Data (Mermilliod, Mermilliod, \& Hauck 1997), 
and 2MASS (Cutri et al. 2003). For comparison with our uncolor-corrected, 
measured fluxes, we list the predicted 24 $\mu$m and 70 $\mu$m fluxes 
integrated over the MIPS bandpasses in Table 2.

\section{Excess Statistics}
Fourteen objects in our sample ($>$35\%) possess 24 $\mu$m fluxes $\geq$30\% 
larger than their predicted photospheres. Seven of which (HD 106906, 
HD 113556, HD 113766, HD 114082, HD 115600, HD 117214, HD 152404) are 
detected at 70 $\mu$m and possess 70 $\mu$m excesses $\geq$100$\times$ larger
than their predicted photospheres. The 24 $\mu$m and 70 $\mu$m fluxes
are probably not interstellar because the point source, at the 
\emph{Hipparcos} positions of the stars, are detected with good contrast to 
the sky. With the exception of AK Sco (HD 152404), the 24 $\mu$m plus 
70 $\mu$m excess sources are members of Lower Centaurus Crux with estimated 
ages $\sim$16 Myr. Seven of our 24 $\mu$m excess objects (HD 103234, 
HD 103703, HD 104231, HD 111102, HD 119511, HD 133075, HD 148040) are not 
detected at 70 $\mu$m. Five of these objects (HD 103234, HD 103703, 
HD 104231, HD 111102, HD 119511) are members of Lower Centaurus Crux, one is 
a member of Upper Centaurus Lupus (HD 133075), and one is a member of Upper 
Scorpius (HD 148040). We infer association subgroup disk fractions of 46\%, 
9\%, and 20\% for each of the Scorpius-Centaurus subgroups respectively from 
our 24 $\mu$m excess sources. Since our sample was selected on the basis
of common proper motion, $\sim$30\% of our sample may be interlopers, 
suggesting that the the Sco-Cen disk fractions measured in this study may be 
as much as 40\% higher than quoted above. More accurate measurements of the 
disk fractions will be made once all of the scheduled \emph{Spitzer} MIPS 
observations are completed. The existence of so many 24 $\mu$m excess sources 
may not be surprising. A $JHKL$ search for near-infrared excesses in the 5-8 
Myr $\eta$ Cha cluster revealed hot inner disks around 60\% (9 out of the 15) 
of the sources searched (Lyo et al. 2003). 

For each 24 $\mu$m plus 70 $\mu$m excess source, we fit the MIPS 24 $\mu$m 
and 70 $\mu$m excess fluxes with a single temperature black body, $T_{gr}$, 
(see Figure 2), and infer color temperatures, $T_{gr}$ = 65 - 330 K, and 
fractional infrared luminosities, $L_{IR}/L_{*}$ = 7 $\times$ 10$^{-4}$ - 
3 $\times$ 10$^{-3}$. Fitting a simple black body to broad-band photometry 
often underestimates the dust grain temperature because the real grain 
emissivity falls more rapidly at wavelengths $\lambda$ $>$ $2 \pi a$. We 
additionally fit the 24 $\mu$m and 70 $\mu$m photometry assuming an emissivity 
$Q_{\lambda}$ = constant for $\lambda$ $<$ $2 \pi a$ and $Q_{\lambda}$ = 
constant ($2 \pi a$/$\lambda$)$^{1.5}$ for $\lambda$ $>$ $2 \pi a$ (Backman \&
Paresce 1993). For HD 113766, this emissivity law is unable to reproduce 
simultaneously the IRAS 12 $\mu$m, the MIPS 24 $\mu$m, and the MIPS 70 
$\mu$m fluxes. If the MIPS 24 $\mu$m and 70 $\mu$m fluxes are used to 
constrain the SED, this model predicts a 12 $\mu$m flux which is significantly 
higher than observed while the simple black body is able to reproduce both the 
MIPS and IRAS data sets. To be consistent, we have used a simple black body to 
model all of the objects in our sample. Recent \emph{Spitzer} IRS spectra have 
revealed the lack of 10 $\mu$m and 20 $\mu$m silicate features around debris 
disks, suggesting that the grains in these systems have diameters larger than 
10 $\mu$m (Jura et al. 2004). We can not constrain the color temperatures of 
the 24 $\mu$m excess only sources without infrared excess detections at 
another wavelength. We infer the fractional dust luminosities for 24 $\mu$m 
excess only sources assuming $F_{IR}$ $\approx$ $\nu F_{\nu}$(24 $\mu$m). Our 
constraint on the 70 $\mu$m fluxes of the 24 $\mu$m excess-only sources 
suggests that the color temperatures for these sources is significantly 
warmer. These sources may be circumstellar disks with warmer dust grains or 
they may be entirely different objects altogether.


\section{Optically Thin Disks?}
One possibility is that these dusty disks are optically thin. The high 
fractional infrared luminosities, $L_{IR}/L_{*}$ = 7 $\times$ 10$^{-4}$ -
3 $\times$10$^{-3}$, associated with stars with ages $\sim$16 Myr, suggests 
that these objects may be similar to $\beta$ Pictoris which possesses 
$L_{IR}/L_{*}$ = 10$^{-3}$ at an age of 12 Myr. The observed 24 $\mu$m and 
70 $\mu$m infrared excesses can be modeled using an optically thin dust 
distribution as described below.

\subsection{Circumstellar Dust Grain Size}
A lower limit to the size of dust grains orbiting a star can be found by
balancing the force due to radiation pressure with the force due to 
gravity. For small grains with radius $a$, the force due to radiation 
pressure overcomes gravity for:
\begin{equation}
a < 3 L_{*} Q_{pr}/(16 \pi G M_{*} c \rho_{s})
\end{equation}
(Artymowicz 1988) where $Q_{pr}$ is the radiation pressure coupling 
coefficient and $\rho_{s}$ is the density of an individual grain. Since 
radiation from F- and G-type stars is dominated by ultraviolet and visual 
light, we expect that $2 \pi a/\lambda \gg 1$ and therefore the effective 
cross section of the grains can be approximated by their geometric cross 
section so $Q_{pr} \approx 1$. Based upon $T_{*}$ and $L_{*}$, the inferred 
stellar masses are 1.4 - 1.8 $M_{\sun}$ (Siess, Dufour, \& Forestini 2000).
With $\rho_{s}$ = 2.5 g cm$^{-3}$, the minimum radii for grains orbiting 
our 24 $\mu$m plus 70 $\mu$m excess objects are $a$ = 0.5 - 1.9 $\mu$m.

We can estimate the average size of the grains assuming a size distribution
for the dust grains. As expected from equilibrium between production and 
destruction of objects through collisions (Greenberg \& Nolan 1989), we assume
\begin{equation}
n(a) da = n_{o} a^{-p} da
\end{equation}
with p $\simeq$ 3.5 (Binzel, Hanner, \& Steel 2000). If we assume a minimum 
grain radius of 0.5 - 1.9  $\mu$m, we find an average grain radius $<a>$ = 
0.8 - 3.3 $\mu$m, if we weight by the number of particles. 

\subsection{Mass of Circumstellar Dust}
The characteristic grain distance can be constrained from the temperature of 
the grains assuming that they are black bodies. We estimate grain temperatures
of $T_{gr}$ = 65 - 330 K from the ratio of the 24 $\mu$m excesses to
70 $\mu$m excesses. Black bodies in radiative equilibrium with a stellar 
source are located a distance
\begin{equation}
D = \frac{1}{2} (\frac{T_{*}}{T_{gr}})^{2} R_{*}
\end{equation}
from the central star (Jura et al. 1998), where $T_{*}$ and $R_{*}$ are the 
effective temperature of the stellar photosphere and the stellar radius. We 
use stellar luminosities, $L_{*}$, from Sartori et al. (2003) and our 
inferred stellar effective temperatures, $T_{*}$, with stellar evolution 
models (Siess, Dufour, \& Forestini 2000) to infer stellar radii, $R_{*}$, 
and masses, $M_{*}$, where possible. For objects with $L_{*}$ and $T_{*}$ 
which do not appear on model grids, we infer stellar radii from $L_{*}$ 
= $4 \pi R_{*}^2 \sigma T_{*}^{4}$. From equation (1) and the stellar 
properties summarized in Table 3, we find characteristic grain distances of 
2.9 AU - 40 AU. 

We can estimate the minimum mass of dust assuming that the particles have 
$<a>$ $\sim$ 0.8 - 3.3 $\mu$m; if the grains are larger, then our estimate is 
a lower bound. If we assume a thin shell of dust at distance, $D$, from the 
star and if the particles are spheres of radius, $a$, and if the cross 
section of the particles equals their geometric cross section, then the mass 
of dust is
\begin{equation}
M_{d} \geq \frac{16}{3} \pi \frac{L_{IR}}{L_{*}} \rho_{s} D^{2} <a> 
\end{equation}
(Jura et al. 1995) 
where $L_{IR}$ is the luminosity of the dust. From equation (3) and the
stellar properties summarized in Table 3, we infer dust masses, $M_{dust}$
= 7 $\times$10$^{-4}$ - 2 $\times$ 10$^{-2}$ $M_{moon}$.

\subsection{Stellar Wind Drag}
Stellar wind drag effectively removes small dust particles around young F- 
and G-type stars. We can compare the mass loss rate from stellar wind drag to 
that of the Poynting-Robertson effect, the dominant grain removal mechanism 
in debris disks around main sequence A-type stars and in our Solar System. 
The increase in ``drag'' in the inward drift velocity over that produced
by the Poytning-Robertson effect is given approximately by the factor
(1 + ${\dot M_{wind}} c^{2}/L_{*}$), where ${\dot M_{wind}}$ is the
stellar wind mass loss rate and $L_{*}$ (Jura 2004). For our solar system, 
${\dot M_{wind}}$ = 2$\times$10$^{12}$ g/sec or 
${\dot M_{wind}} c^{2}/L_{\sun}$ = 0.5. 

We infer ${\dot M_{wind}}$ from \emph{ROSAT} fluxes for our sample using the 
observed dependence of stellar mass loss rate, ${\dot M_{wind}}$, per stellar 
surface area, $A$, on x-ray flux per stellar area for nearby G-, K- and, 
M-type stars ${\dot M_{wind}}/A \propto F_{x}^{1.15\pm0.2}$ scaled to 
observations of 36 Oph ($F_{x}$ = 3.6 $\times$ 10$^{5}$ erg cm$^{-2}$ 
s$^{-1}$, ${\dot M_{wind}}/A$ = 17 ${\dot M_{\sun}}/A_{\sun}$; Wood et al. 
2002). Using the \emph{ROSAT} fluxes listed in Table 1, we estimate 
${\dot M_{wind}} c^{2}/L_{\sun}$ = 76, 27, 15 and 77 for HD 103234, HD 104231,
HD 113766, and HD 148040 respectively, suggesting that stellar wind drag is 
more important than the Poynting-Robertson effect for all of these systems. 
The nondetection of the remaining sources in the \emph{ROSAT} All-Sky Bright 
Source and the \emph{ROSAT} All-Sky Survey Faint Source Catalogs suggests 
that they possess 0.2 - 2.0 keV fluxes $<$0.05 counts/sec. For these objects, 
we can not place strong constraints on ${\dot M_{wind}} c^{2}/L_{\sun}$ to 
determine the relative importance of stellar wind drag and the 
Poynting-Robertson effect; however, the stellar wind mass loss rates are 
expected to be similar to the other stars in the sample. Solar evolutionary 
models allow for a Sun which, at an age of 20 Myr, possesses a mass loss rate 
as high as, ${\dot M_{wind}}$ = 2 ${\times}$ 10$^{15}$ g s$^{-1}$ (Sackmann 
\& Boothroyd 2003), suggesting ${\dot M_{wind}} c^{2}/L_{\sun}$ = 460.   

\subsection{Lifetime of Circumstellar Grains}
We can estimate the lifetime of circumstellar dust grains assuming that
stellar wind drag is the dominant removal mechanism. The speed at which 
particles drift radially inward under the Poynting-Robertson effect and 
stellar wind drag is
\begin{equation}
v = \left( \frac{3 L_{*}}{8 \pi a \rho_{s} c^2 D} \right)
    \left( 1 + \frac{{\dot M_{wind}} c^{2}}{L_{*}} \right)
\end{equation}
(Burns, Lamy, \& Soter 1979; Gustafson 1994); therefore, the lifetime for 
particles under stellar wind drag, $t_{drag}$, in environments in which
stellar wind drag is much larger than the Poynting-Robertson effect 
(${\dot M_{wind}} c^{2}/L_{*}$ $>>$ 1) is:
\begin{equation}
t_{drag} = \frac{4 \pi <a> \rho_{s} D^{2}}{3 {\dot M_{wind}}}
\end{equation}
For HD 113766, if $<a>$ = 3.3 $\mu$m, $\rho_{s}$ = 2.5 g cm$^{-3}$, 
$D$ = 2.9 AU, ${\dot M_{wind}}$ = 760 ${\dot M_{\sun}}$, then
$t_{drag}$ = 12,000 yr. Since this timescale is significantly shorter than
the stellar age ($t_{age}$), we hypothesize that the grains are replenished
by collisions between larger bodies. This system may be a younger, lower
mass analog to $\zeta$ Lep, a 200 Myr A3V member of the Castor Moving Group, 
located 19 pc away from the Sun (Barrado y Navascu\'{e}s 1998) which 
possesses dust at terrestrial temperatures, T$_{gr}$ $\sim$ 300 K. Since the 
Poynting-Robertson drag lifetime of dust grain in this system is 10,000 
years, the dust grains must be replenished by collisions between asteroidal 
bodies. High resolution 11.7 $\mu$m and 17.9 $\mu$m imaging and photometry 
confirmed that the dust grains are are located at a distance $\leq$6 AU from 
the star (Chen \& Jura 2001).  

\subsection{Minimum Mass of the Parent Bodies}
We can estimate the minimum mass in parent bodies assuming that the system is 
in a steady state and that the minimum mass is larger than the mass already 
removed by the stellar wind during the lifetime of the star. If $M_{PB}$ 
denotes the minimum parent body mass, then we may write
\begin{equation}
M_{PB} \geq {\dot M_{dust}} t_{age}
\end{equation}
where ${\dot M_{dust}}$ is the dust removal rate
\begin{equation}
{\dot M_{dust}} = {\dot M_{wind}} \frac{L_{IR}}{L_{*}}
\end{equation}
Jura (2004). Using equation (8), stellar wind mass loss rates inferred
from \emph{ROSAT} fluxes given in Table 1 and fractional infrared 
luminosities listed in Table 3, we infer ${\dot M_{dust}}$ $<$ 2 
$\times$ 10$^{-6}$ $M_{moon}$/yr for all of the 24 $\mu$m excess systems. 
Using equations (7) and (8), we estimate:
\begin{equation}
M_{PB} \geq \frac{L_{IR}}{L_{*}} {\dot M_{wind}} t_{age}
\end{equation}
From equation (9) and the stellar properties listed in Tables (1) and (3),
we infer parent body masses $M_{PB}$ = 0.3 - 10 $M_{moon}$, approximately
10 times the mass of the Main Asteroid Belt in our Solar System. The largest
minimum parent body mass, 10 $M_{moon}$, is close to 0.2 $M_{\earth}$,
suggesting that these systems may possess environments in which planets
may be forming.


\section{Flat, Optically Thick Disks?}
The candidate disks, discovered in this survey, have approximately the same
70 $\mu$m luminosity. One natural explanation for the convergence of the
70 $\mu$m luminosities is that the disks are optically thick and 
geometrically thin, similar to the disk found around the 10 Myr old TW Hydrae 
Association object HD 98800. Face-on, opaque, geometrically thin disks with 
inner holes can be constructed so that they possess color temperatures 
consistent with the MIPS data. 

The predicted temperature structure for a geometrically flat, passive, 
optically thick circumstellar disk is
\begin{equation}
T_{disk} = \left( \frac{2}{3 \pi} \right)^{1/4}
            \left( \frac{R_{*}}{R} \right)^{3/4} T_{*}
\end{equation}
(Jura 2003). The predicted flux density for the same disk with an 
inclination, $i$, at a frequency, $\nu$, is
\begin{equation}
F_{\nu} = 12 \pi^{1/3} \frac{R_{*}^2 \cos i}{D_{*}^2}
    \left( \frac{2 k_{B} T_{*}}{3 h \nu} \right)^{8/3}
    \frac{h \nu^{3}}{c^{2}} \int_{x_{in}}^{x_{out}} \frac{x^{5/3}}{e^{x}-1} dx
\end{equation}
(Jura 2003) where $x$ = $h \nu/k_{B} T_{disk}$ and $D_{*}$ is the distance to 
the star. We use the measured $F_{\nu}$(24 $\mu$m)/$F_{\nu}$(70 $\mu$m) flux 
ratio to constrain the inner radii for the disks, assuming that they have 
$R_{out}$ = $\infty$, and the magnitude of the 24 $\mu$m flux to infer the 
disk inclinations. For example, for HD 106906 with $D_{*}$ = 92 pc, 
$T_{*}$ = 6750 K, and $R_{*}$ = 1.6 $R_{\sun}$, we estimate an inner 
radius, $R_{in}$ = 0.68 AU and an inclination, $i$ = 68 $\arcdeg$. The 
measured fluxes are approximately a factor of 2 smaller than the maximum
value allowed for an opaque, flat disk. The fact that (1) the measured fluxes
are so close to the maximum allowed value and (2) the estimated inclination 
angles appear to cluster around 65$\arcdeg$ may suggest saturation in our
model. The saturation may be the result of our simple grain assumptions.
We have not considered the possibility that circumstellar dust grains might 
reflect some of the incident stellar radiation as observed in some 
circumstellar disks. If the albedo in these systems is $\sim$0.5, then the
predicted 24 $\mu$m and 70 $\mu$m fluxes would be reduced by a factor of 2.
Spatially resolved scattered light imaging, combined with thermal mid-infrared 
imaging of the HD 141569, suggest that the albedo in this system is $\sim$0.4, 
similar to the values inferred for Herbig Ae/Be stars using \emph{ISO} 
(Weinberger et al. 1999). Our inferred disk inner radii are similar to those 
inferred for T-Tauri disks. Near-infrared spectroscopy suggests that the inner 
radii of T-Tauri disks are located at the dust sublimation radii, $R_{in}$ = 
0.07-0.5 AU, further than expected if the dust is heated by the luminosity of 
the star alone because the disk is heated by both the stellar luminosity and 
accretion onto the star (Muzzerolle et al. 2003). Whether these systems are 
continuing to accrete gas is not obervationally known. 

One star which does not fit the model of an optically thick, geometrically 
thin disk with an inner hole is HD 113766 which possess a significantly 
stronger 24 $\mu$m excess than a 70 $\mu$m excess. For HD 113766, we use the 
measured $F_{\nu}$(2.2 $\mu$m)/$F_{\nu}$(24 $\mu$m) and the measured
$F_{\nu}$(24 $\mu$m)/$F_{\nu}$(70 $\mu$m) flux ratios to constrain the 
inner and outer radii for the disk; however, we were unable to scale
the magnitude of the 24 $\mu$m flux to estimate the disk inclination. The 
dusty disk around HD 113766 produces more radiation than can be produced by 
an optically thick, geometrically thin disk. 

\section{Discussion}
The current data do not allow us to determine whether these dusty disks
possess grains at a single radius or whether they possess continuous disks
(for example, disks whose surface density distributions are determined by 
infall under stellar wind drag), or whether they possess flat, optically 
thick disks. 

Although coronal x-ray emission is used as an indicator for youth among
later-type stars, it may not be correlated with the presence of infrared 
excess indicative of the presence of circumstellar dust. In Figure 3, we show 
the observed x-ray luminosities ($L_{x}/L_{bol}$), inferred from \emph{ROSAT} 
observations using the conversion 1 \emph{ROSAT} count = (8.31 + 
5.30$\times$HR1) 10$^{-12}$ erg cm$^{-2}$ (Fleming et al. 1995), where 
HR1 is the hardness ratio between the 0.1 - 0.4 keV and the 0.5 - 2.0 keV 
bands, plotted as a function of infrared luminosity ($L_{IR}/L_{bol}$), 
inferred from MIPS 24 $\mu$m excesses using $L_{IR}$ = 
$\nu L_{\nu}$(24 $\mu$m). Contrary to expectation, the infrared luminosity 
appears anticorrelated with x-ray luminosity expect for 30\% of the objects 
which possess neither a \emph{ROSAT} flux nor a MIPS 24 $\mu$m excess. The
anticorrelation can be naturally explained if stellar wind drag effectively
removes dust grains around young stars with high x-ray coronal activity. The 
objects that lack both a \emph{ROSAT} flux and a MIPS 24 excess may be the 
$\sim$30\% interlopers predicted by de Zeeuw et al. (1999). Recent 
\emph{Chandra} observations of HD 98800, a quadruple system in the 
$\sim$10 Myr old TW Hydrae Association, have revealed that the x-ray flux of 
the dusty binary system HD 98800B is $\sim$4 times fainter than its dustless 
companion HD 98800A (Kastner et al. 2004). Since the x-ray luminosities of 
solar-like stars vary with time, more observations of HD 98800 are needed to 
determine whether the measured anticorrelation between infrared luminosity 
and x-ray luminosity is an effect of observing the system at a special time.

Recent ground-based searches for new 10 $\mu$m and 20 $\mu$m excesses around 
proper motion and x-ray selected K- and M-type members of the $\sim$10 Myr 
old TW Hydrae Association (Weinberger et al. 2004) and around proper motion, 
x-ray and lithium selected F-, G-, K-, and M-type members of the $\sim$30 Myr 
old Tucana-Horologium Association (Mamajek et al. 2004) have been relatively 
unsuccesful. The observed anti-correlation between x-ray luminosity and 
infrared luminosity may help explain the lack of infrared excess sources 
found in the TW Hydrae and Tucana-Horologium Assoications.

\section{Conclusions}

We have obtained MIPS 24 $\mu$m and 70 $\mu$m photometry of 40 F- and G-type 
stars in Scorpius-Centaurus using MIPS on the \emph{Spitzer Space Telescope}.

1. Fourteen targets possess 24 $\mu$m excesses, corresponding to a 35\% disk 
fraction in Scorpius-Centaurus. If the interlopers in the study do not
possess infrared excess and are excluded, then the disk fraction may be as 
high as 50\%. Half of the 24 $\mu$m excess sources also possess 70 $\mu$m 
fluxes $>$100$\times$ larger than expected from their photospheres alone.

2. The sources with 24 $\mu$m plus 70 $\mu$m excesses possess grain 
temperatures, $T_{gr}$ = 65 - 310 K and high fractional infrared luminosities 
$L_{IR}/L_{*}$ = 7.0 $\times$10$^{-4}$ - 0.019 suggesting that these systems 
may possess optically thin, $\beta$ Pictoris-like dusty disks or compact,
optically thick HD 98800-like disks.


3. The observed fractional infrared luminosities of dusty disks appear 
anticorrelated with \emph{ROSAT} x-ray fluxes and may help explain the
difficulty of finding dusty disks around stars which were identified to
be young using x-ray observations.

\acknowledgements
We would like to thank V. Krause, J. Stansberry, and K. Su for their 
assistance with the MIPS DAT and/or the photosphere fitting tools. We would 
like to thank G. Bryden, E. Mamajek, and M. Werner for their comments and 
suggestions. This work is based on observations made with the \emph{Spitzer 
Space Telescope}, which is operated by the Jet Propulsion Laboratory, 
California Institute of Technology under NASA contract 1407.

\begin{deluxetable}{rcccccccc}
\singlespace
\tablecaption{Stellar Properties} 
\tablehead{
    \colhead{HD} &
    \colhead{Spectral} &
    \colhead{Distance} &
    \colhead{$A_{V}$} &
    \colhead{\emph{ROSAT}} &
    \colhead{${\dot M_{wind}}$} &
    \colhead{$t_{age}$} &
    \colhead{Association }&
    \colhead{Notes} \\
    \omit &
    \colhead{Type} &
    \colhead{(pc)} &
    \colhead{(mag)} &
    \colhead{(counts/sec)} &
    \colhead{(${\dot M_{\sun}}$)} &
    \colhead{(Myr)} &
    \omit &
    \omit \\
}

\tablewidth{0pt}
\tablecolumns{9}
\startdata
    98660  &  F2V   &  86 & 0.08 & 0.0924 &     790 & $\sim$16 & LCC \\
    103234 & F2IV   & 105 & 0.06 & 0.0704 &    1200 & $\sim$16 & LCC \\
    103599 & F3IV   &  91 & 0.05 & 0.0395 &     440 & $\sim$16 & LCC \\
    103703 &  F3V   & 105 & 0.09 & ...    &  $<$800 & $\sim$16 & LCC \\
    104231 &  F5V   &  89 & 0.07 & 0.0409 &     240 & $\sim$16 & LCC \\
    104897 &  F3V   & 114 & 0.03 & ...    &  $<$750 & $\sim$16 & LCC \\
    105070 &  G1V   & 102 & 0.00 & 0.219  &    4600 &    13    & LCC \\
    106218 &  F2V   & 107 & 0.10 & 0.0303 &     640 & $\sim$16 & LCC \\
    106444 &  F5V   & 101 & 0.05 & 0.565  &   12000 & $\sim$16 & LCC \\
    106906 &  F5V   &  92 & 0.02 & ...    &  $<$600 & $\sim$16 & LCC \\
    107920 &  F3V   & 126 & 0.03 & ...    & $<$1200 & $\sim$16 & LCC \\
    108568 &  G1    & 142 & 0.00 & 0.168  &    6500 &    14    & LCC \\
    108611 &  G5V   &  95 & 0.00 & 0.0421 &     420 &    10    & LCC \\
    111102 &  F0III & 120 & 0.00 & ...    & $<$1100 & $\sim$16 & LCC \\
    111104 &  F0    & 146 & 0.11 & 0.145  &    4900 & $\sim$16 & LCC \\
    112509 &  F3IV  & 111 & 0.00 & 0.0265 &     450 & $\sim$16 & LCC \\
    113376 &  G3V   &  94 & 0.05 & ...    &  $<$690 &     ?    & ... \\
    113556 &  F2V   & 102 & 0.06 & ...    &  $<$750 & $\sim$16 & LCC \\
    113766 &  F3    & 131 & 0.01 & 0.0343 &     760 & $\sim$16 & LCC & binary \\
    114082 &  F3V   &  83 & 0.09 & ...    &  $<$470 & $\sim$16 & LCC \\
    115600 &  F2IV  & 111 & 0.04 & ...    &  $<$910 & $\sim$16 & LCC \\
    117214 &  F6V   &  97 & 0.04 & ...    &  $<$690 & $\sim$16 & LCC \\
    117945 &  F7    &  92 & 0.06 & ...    &  $<$620 & $\sim$16 & LCC \\
    119022 &  G2IV  & 124 & 0.11 & ...    & $<$1300 & $\sim$16 & LCC \\
    119511 &  F3V   &  91 & 0.01 & ...    &  $<$580 & $\sim$16 & LCC \\
    125912 &  F7V   & 115 & 0.12 & 0.0819 &    2000 & $\sim$17 & UCL \\
    133075 &  F3IV  & 137 & 0.32 & ...    & $<$1500 & $\sim$17 & UCL \\
    136991 &  F3V   & 114 & 0.01 & ...    &  $<$970 & $\sim$17 & UCL \\
    137130 &  F0V   & 136 & 0.16 & ...    & $<$1400 & $\sim$17 & UCL \\
    138009 &  G6V   &  92 & 0.00 & 0.609  &   12000 &     7    & UCL \\
    138398 &  G6III & 146 & 0.00 & ...    & $<$1900 & $\sim$17 & UCL \\
    138994 &  F2V   & 149 & 0.04 & ...    & $<$1800 & $\sim$17 & UCL \\
    139883 &  F2V   &  97 & 0.04 & ...    &  $<$670 & $\sim$17 & UCL \\
    142113 &  F8V   &  90 & 0.34 & 0.119  &    2000 & $\sim$5  & US  \\
    143677 &  G8V   & 143 & 0.03 & 0.409  &   19000 & $\sim$17 & UCL \\
    144225 &  F3V   & 119 & 0.26 & 0.320  &    8000 & $\sim$17 & UCL \\
    145208 &  G5V   & 121 & 0.08 & 0.250  &   11000 & $\sim$5  & US  \\
    148040 &  G0V   & 121 & 0.00 & 0.257  &    8700 & $\sim$5  & US  \\
    148153 &  F5V   & 128 & 0.03 & ...    & $<$1300 & $\sim$5  & US  \\
    152057 &  F0V   & 150 & 0.13 & ...    & $<$1800 & $\sim$5  & US  \\
    152404 &  F5V   & 145 & 0.75 & ...    & $<$1700 & $\sim$17 & UCL \\
\enddata
\tablecomments{LCC = Lower Centaurus Crux, US = Upper Scorpius,
    UCL = Upper Centaurus Lupus}
\end{deluxetable}


\begin{deluxetable}{rccccccc}
\singlespace
\tablecaption{Uncolor-Corrected MIPS 24 $\mu m$ and 70 $\mu m$ Fluxes} 
\tablehead{
    \omit &
    \omit &
    \colhead{Measured} &
    \colhead{Predicted} &
    \colhead{Measured} &
    \colhead{Measured} &
    \colhead{Predicted} &
    \colhead{COBE} \\
    \omit &
    \omit &
    \colhead{MIPS} &
    \colhead{Photosphere} &
    \colhead{MIPS} &
    \colhead{70 $\mu$m} &
    \colhead{Photosphere} &
    \colhead{70 $\mu$m} \\
    \colhead{HD} &
    \colhead{AOR} &
    \colhead{F$_{\nu}$(24 $\mu$m)} &
    \colhead{F$_{\nu}$(24 $\mu$m)} &
    \colhead{F$_{\nu}$(70 $\mu$m)} &
    \colhead{SNR} &
    \colhead{F$_{\nu}$(70 $\mu$m)} &
    \colhead{Background} \\
    \omit &
    \colhead{ID} &
    \colhead{(mJy)} &
    \colhead{(mJy)} &
    \colhead{(mJy)} &
    \omit &
    \colhead{(mJy)} &
    \colhead{(MJy sr$^{-1}$)} \\
}
\tablewidth{0pt}
\tablecolumns{6}
\startdata
    98660   & 4778752 &   10.8 & 10.9 & $<$33 & ... & 1.2 &  9.8 \\
    103234  & 4780032 &   19.3 &  9.0 & $<$24 & ... & 1.0 &  18.8 \\
    103599  & 4780544 &    9.6 &  9.6 & $<$28 & ... & 1.1 &  13.2 \\
    103703  & 4780800 &   26.7 &  8.0 & $<$33 & ... & 0.9 &  21.8 \\
    104231  & 4781056 &   18.4 &  7.9 & $<$33 & ... & 0.9 &  17.5 \\
    104897  & 4781312 &    9.2 &  8.3 & $<$24 & ... & 0.9 &  10.9 \\
    105070  & 4781568 &    9.4 &  9.2 & $<$23 & ... & 1.0 &  11.5 \\
    106218  & 4782336 &    7.7 &  7.7 & $<$21 & ... & 0.9 &  16.9 \\
    106444  & 4782848 &   10.1 &  9.1 & $<$25 & ... & 1.0 &  15.9 \\
    106906  & 4783872 &  108   & 15.3 &   290 &  38 & 1.7 &  16.1 \\
    107920  & 4784896 &    6.0 &  7.5 & $<$30 & ... & 0.9 &  21.8 \\
    108568  & 4785152 &    9.4 &  9.5 & $<$26 & ... & 1.1 &  16.9 \\
    108611  & 4785408 &    9.8 & 10.7 & $<$41 & ... & 1.2 &  30.4 \\
    111102  & 4786944 &   34.1 & 18.5 & $<$43 & ... & 2.1 &  24.4 \\
    111104  & 4787200 &    9.8 & 10.3 & $<$35 & ... & 1.1 &  24.4 \\
    112509  & 4787968 &    7.9 &  7.7 & $<$39 & ... & 0.9 &  15.5 \\
    113376  & 4788736 &   14.3 & 19.1 & $<$43 & ... & 2.1 &  31.2 \\
    113556  & 4789504 &   19.2 &  9.4 &   170 &  14 & 1.1 &  24.4 \\
    113766  & 4789760 & 1470   & 18.6 &   350 &  44 & 2.1 &  10.5 \\
    114082  & 4790784 &  223   & 11.3 &   340 &  15 & 1.3 &  40.1 \\
    115600  & 4791552 &  122   &  8.6 &   160 &   9 & 1.0 &  31.1 \\
    117214  & 4792320 &  207   & 11.9 &   300 &  17 & 1.3 &  26.9 \\
    117945  & 4792832 &    9.5 & 12.3 & $<$52 & ... & 1.4 &  29.0 \\
    119022  & 4793344 &   39.3 & 40.8 & $<$25 & ... & 4.6 &  14.9 \\
    119511  & 4793600 &   10.4 &  7.4 & $<$29 & ... & 0.8 &  12.5 \\
    125912  & 4796672 &   17.0 & 16.0 & $<$30 & ... & 1.8 &  12.7 \\
    133075  & 4798720 &   22.6 & 15.2 & $<$28 & ... & 1.7 &  17.8 \\
    136991  & 4800768 &    7.1 &  7.7 & $<$43 & ... & 0.9 &  17.4 \\
    137130  & 4801024 &    7.5 &  8.4 & $<$27 & ... & 0.9 &  20.9 \\
    138009  & 4801536 &   14.4 & 12.5 & $<$22 & ... & 1.4 &  18.1 \\
    138398  & 2802304 &   27.2 & 30.2 & $<$44 & ... & 3.4 &  25.3 \\
    138994  & 4802560 &    9.6 &  9.4 & $<$24 & ... & 1.0 &  17.7 \\
    139883  & 4802816 &    7.6 &  7.8 & $<$27 & ... & 0.9 &  21.2 \\
    142113  & 4804608 &    9.6 &  9.6 & $<$27 & ... & 1.1 &  25.1 \\ 
    143677  & 4805632 &   10.0 & 10.4 & $<$45 & ... & 1.2 &  29.2 \\
    144225  & 4805888 &   14.8 & 15.0 & $<$37 & ... & 1.7 &  22.6 \\
    145208  & 4806656 &   10.8 & 10.9 & $<$28 & ... & 1.2 &  31.5 \\
    148040  & 4809984 &   12.9 &  9.7 & $<$59 & ... & 1.1 &  47.2 \\
    148153  & 4810240 &   12.6 & 10.9 & $<$71 & ... & 1.2 &  51.1 \\
    152057  & 4812032 &    8.9 &  7.9 & $<$28 & ... & 0.9 &  26.0 \\
    152404  & 4812544 & 3390   & 16.8 &  2840 & 120 & 1.2 &  40.1 \\ 
\enddata
\end{deluxetable}


\begin{deluxetable}{lcccccccccc}
\singlespace
\tablecaption{Single Temperature Disk Model Parameters} 
\tablehead{
    \colhead{HD} & 
    \colhead{$T_{*}$} &
    \colhead{$L_{*}$} &
    \colhead{$R_{*}$} &
    \colhead{$M_{*}$} &
    \colhead{$T_{gr}$} &
    \colhead{$L_{IR}/L_{*}$} &
    \colhead{D} &
    \colhead{$M_{dust}$} &
    \colhead{${\dot M_{dust}}$} &
    \colhead{$M_{PB}$} \\
    \omit &
    \colhead{(K)} &
    \colhead{($L_{\sun}$)} &
    \colhead{($R_{\sun}$)} &
    \colhead{($M_{\sun}$)} &
    \colhead{(K)} &
    \omit &
    \colhead{(AU)} &
    \colhead{($M_{moon}$)} &
    \colhead{($M_{moon}/yr$)} &
    \colhead{($M_{moon}$)} \\
}
\tablewidth{0pt}
\tablecolumns{11}
\startdata
103234 & 7000 &  4.6 & 1.4 & 1.4 & ... & 1.0$\times$10$^{-4}$ & ... & 
    ... & 6$\times$10$^{-8}$ & 1 \\ 
103703 & 7000 &  3.9 & 1.4 & ... & ... & 1.9$\times$10$^{-4}$ & ... &
    ... & $<$8$\times$10$^{-8}$ &... \\
104231 & 6750 &  2.6 & 1.2 & ... & ... & 1.4$\times$10$^{-4}$ & ... &
    ... & 2$\times$10$^{-8}$ & 0.3 \\
106906 & 6750 &  5.2 & 1.6 & 1.4 &  90 & 1.4$\times$10$^{-3}$ &  20 &
    1$\times$10$^{-2}$ & $<$5$\times$10$^{-7}$ & ... \\
111102 & 8000 & 19.1 & 2.3 & 1.8 & ... & 4.3$\times$10$^{-5}$ & ... &
    ... & $<$2$\times$10$^{-8}$ & ... \\
113556 & 7250 &  4.8 & 1.4 & 1.5 &  70 & 7.0$\times$10$^{-4}$ &  40 &
    2$\times$10$^{-2}$ & $<$3$\times$10$^{-7}$ & ... \\
113766 & 6750 & 14.1 & 2.6 & 1.8 & 330 & 2.1$\times$10$^{-3}$ &  3 &
    5$\times$10$^{-3}$ & 8$\times$10$^{-7}$ & 10 \\
114082 & 6750 &  3.2 & 1.3 & ... & 110 & 3.0$\times$10$^{-3}$ &  10 &
    5$\times$10$^{-3}$ & $<$8$\times$10$^{-7}$ & ... \\
115600 & 7250 &  5.2 & 1.4 & 1.5 & 120 & 1.6$\times$10$^{-3}$ &  10 &
    5$\times$10$^{-3}$ & $<$8$\times$10$^{-7}$ & ... \\
117214 & 6750 &  4.6 & 1.5 & 1.4 & 110 & 2.4$\times$10$^{-3}$ &  10 &
    7$\times$10$^{-4}$ & $<$9$\times$10$^{-7}$ & ... \\
119511 & 7000 &  2.5 & 1.1 & ... & ... & 3.8$\times$10$^{-5}$ & ... &
    ... & $<$1$\times$10$^{-8}$ & ... \\
133075 & 6750 & 13.5 & 2.5 & 1.8 & ... & 3.5$\times$10$^{-5}$ & ... &
    ... & $<$3$\times$10$^{-8}$ & ... \\
148040 & 6250 &  4.4 & 1.8 & 1.3 & ... & 4.4$\times$10$^{-4}$ & ... &
    ... & 2$\times$10$^{-6}$ & 10 \\
\enddata
\end{deluxetable}


\begin{deluxetable}{lccc}
\singlespace
\tablecaption{Optically Thick Disk Model Parameters} 
\tablehead{
    \colhead{HD} & 
    \colhead{$R_{in}$} &
    \colhead{$R_{out}$} &
    \colhead{$i$} \\
    \omit &
    \colhead{(AU)} &
    \colhead{(AU)} &
    \colhead{($\arcdeg$)} \\
}
\tablewidth{0pt}
\tablecolumns{4}
\startdata
    106906 & 0.68  & $\infty$ & 68 \\
    113556 & 1.4   & $\infty$ & 60 \\
    114082 & 0.32  & $\infty$ & 63 \\
    115600 & 0.32  & $\infty$ & 72 \\
    117214 & 0.34  & $\infty$ & 63 \\
\enddata
\end{deluxetable}

\begin{figure}
\figurenum{1}
\plotone{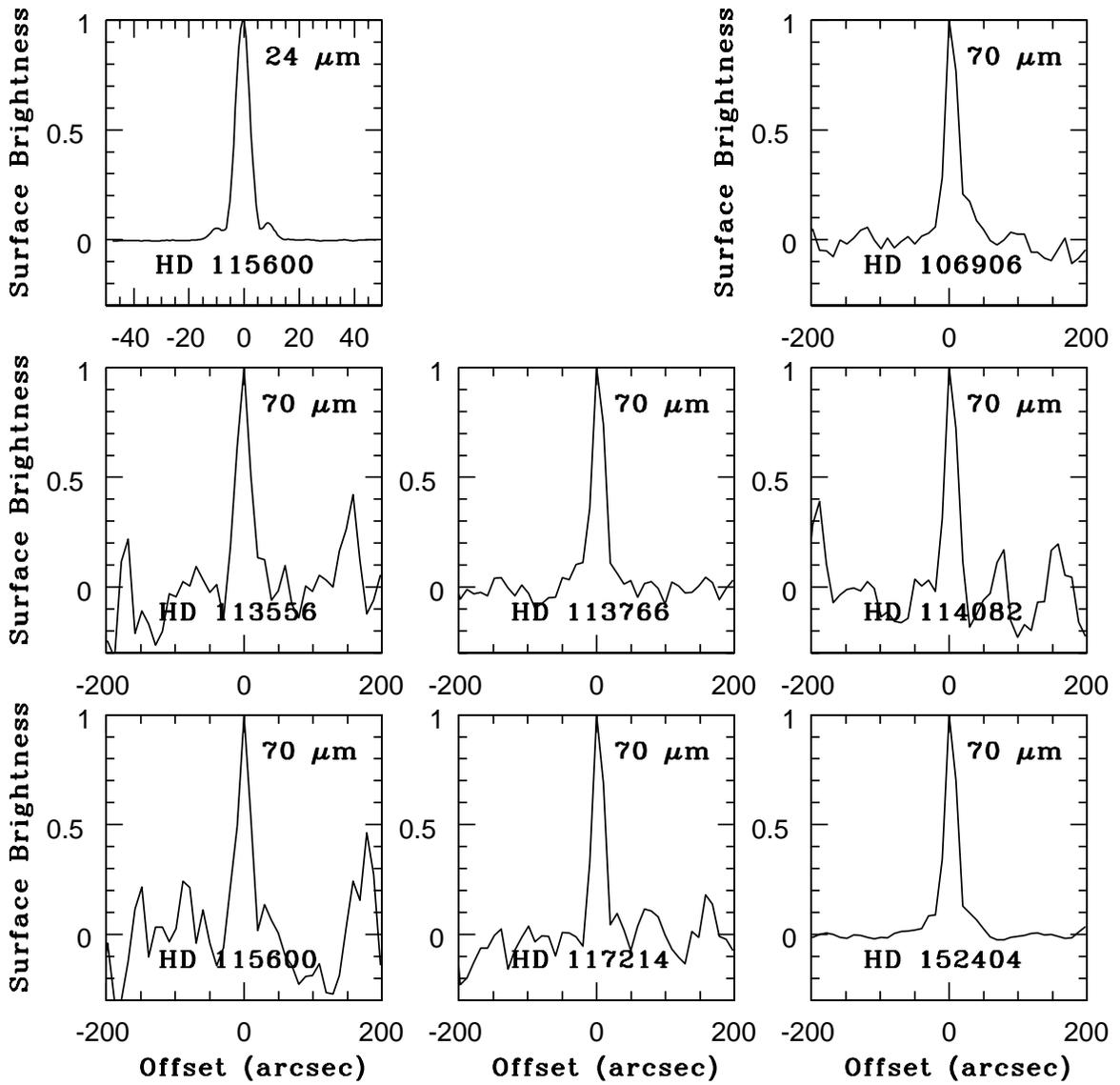}
\caption{Normalized surface brightness line cuts through HD 115600 at 
24 $\mu$m (representative of all MIPS 24 $\mu$m detections in this study)
and through all of the 70 $\mu$m sources at 70 $\mu$m.
All sources appear as point source when detected at either wavelength and 
have positions coincident with \emph{Hipparcos} stellar positions, 
suggesting that the far-infrared emission is circumstellar and not 
interstellar.}
\end{figure}

\begin{figure}
\figurenum{2}
\plotone{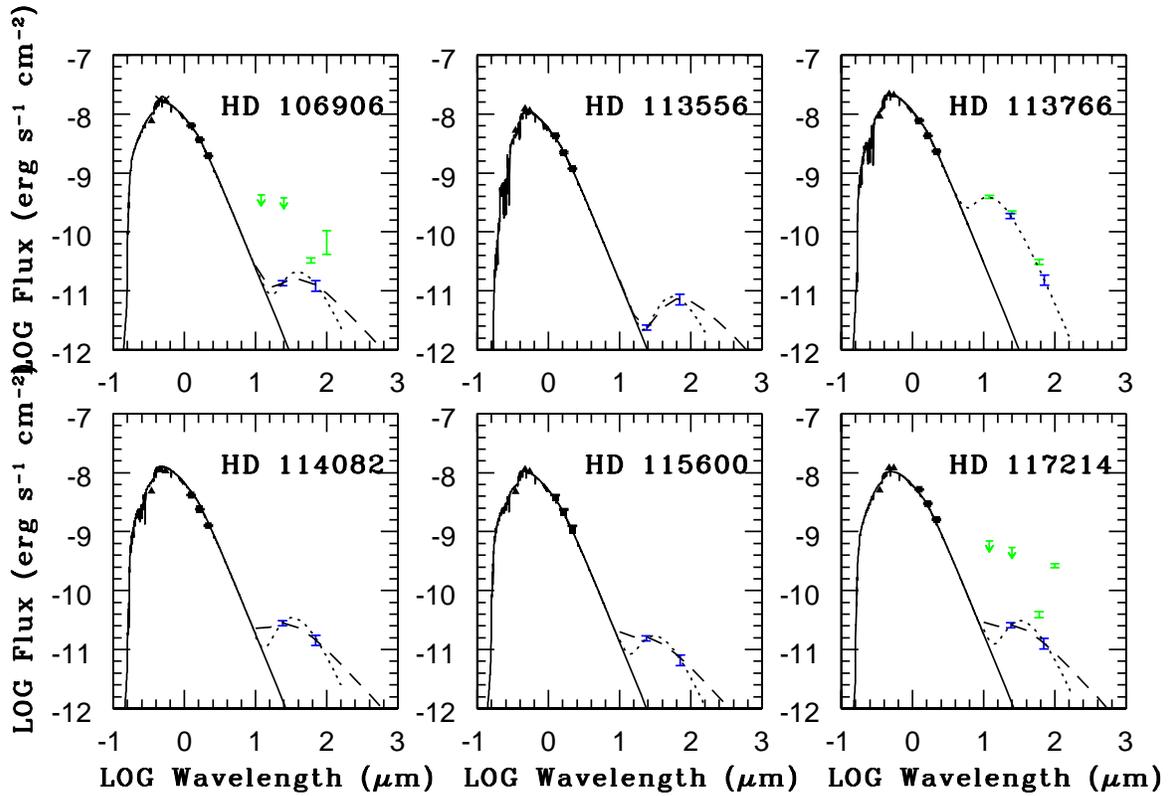}
\caption{Spectral Energy Distributions (SEDs) for objects which possesses 
24 $\mu$m plus 70 $\mu$m excesses: HD 106906, HD 113556, HD 113766, 
HD 114082, HD 115600, HD 117214. GCPD mean $uvby$ fluxes are plotted as 
triangles, GCPD mean UBV fluxes are plotted as crosses, and 2MASS JHK fluxes 
(Cutri et al. 2003) are plotted as squares. \emph{IRAS} photometry, where 
available, is shown with green error bars. Our MIPS 24 $\mu$m and 70 $\mu$m 
photometry, as reported here, is shown with dark blue error bars. Overlaid 
are the best fit 1993 Kurucz model for the stellar atmospheres. The dotted 
lines are a blackbody fits to the MIPS fluxes at wavelengths longer than 
10 $\mu$m. The dashed lines are the optically thick models. For HD 106906
and HD 117214, the IRAS 60 $\mu$m flux appears brighter than expected from
the MIPS 70 $\mu$m flux. These lines-of-sight possess high cirrus backgrounds.
Structure in the interstellar cirrus observed toward these objects may be
better resolved and subtracted in the \emph{Spitzer} data than in the IRAS
data.}
\end{figure}

\begin{figure}
\figurenum{3}
\plotone{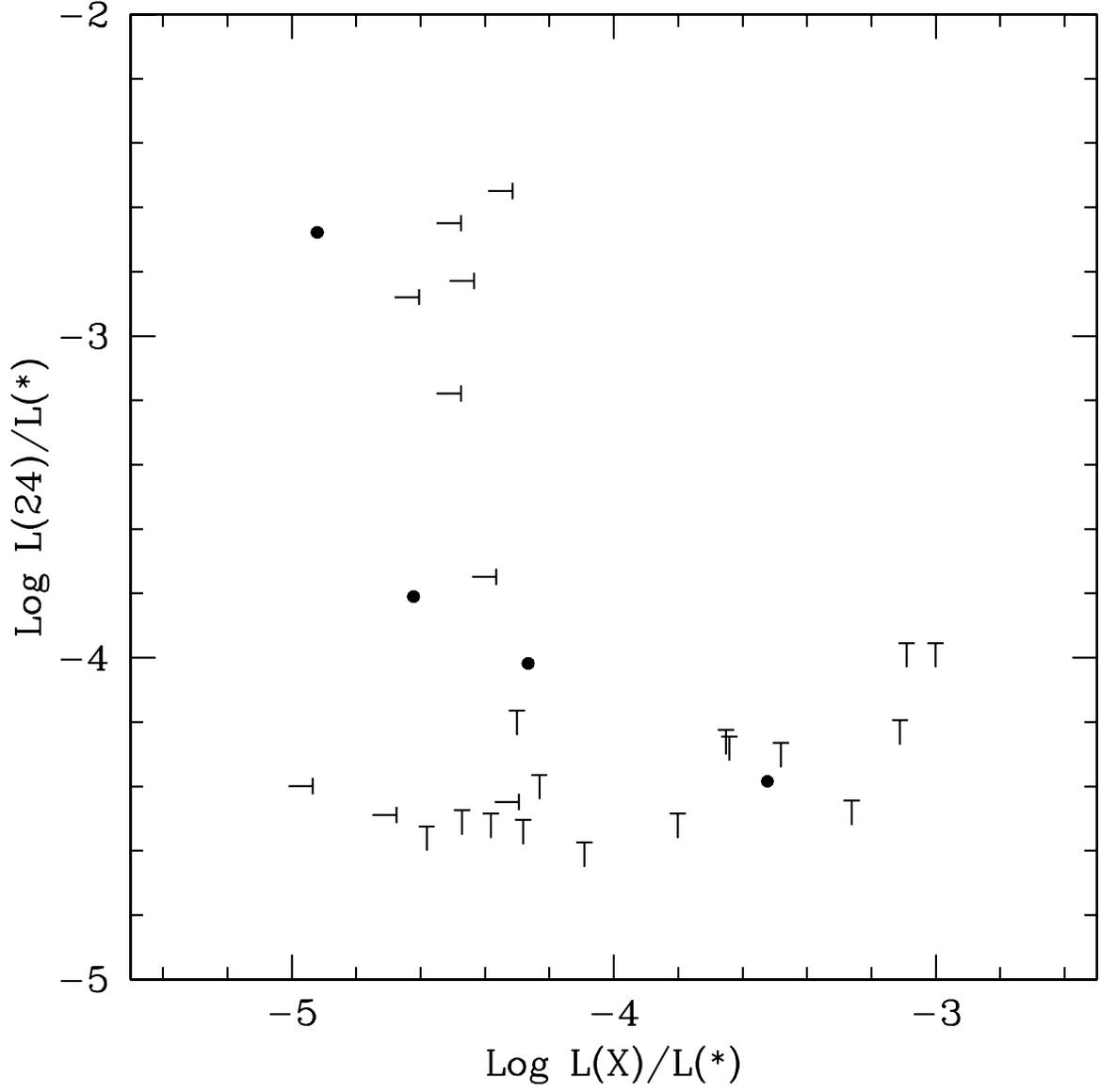}
\caption{Infrared luminosity plotted as a function of stellar x-ray
luminosity. Objects plotted as filled circles have both 24 $\mu$m excesses
and \emph{ROSAT} counterparts. Objects without MIPS 24 $\mu$m excesses
or without \emph{ROSAT} counterparts are plotted with arrows which
show upper limits on the detections.}
\end{figure}

\end{document}